\documentclass[runningheads,a4paper]{llncs}

\usepackage{eso-pic}
\AddToShipoutPicture*{
\put(133,160){
\scriptsize
Proceedings of WLPE 2013, {\tt arXiv:1308.2055}, August 2013.
}}

\usepackage{rotating}
\usepackage{amssymb}
\usepackage{amsmath}
\usepackage{url}
\usepackage[all]{xy}
\usepackage{color}
\usepackage{rotating}
\usepackage{listings}

\usepackage{multicol}
\usepackage{subfig}

\usepackage{cancel}
\makeatletter
\newcommand{\xRightarrow}[2][]{\ext@arrow 0359\Rightarrowfill@{#1}{#2}}
\makeatother

\usepackage{graphics}
\usepackage{graphicx}
\usepackage{wrapfig}
\usepackage{enumitem}

\newcommand{\true}{\texttt{true}}

\newcommand{\plus}{\texttt{int\_plus}}

\newcommand{\set}[1]{\left\{
      \begin{array}{l}\!\!#1\!\!\end{array}
      \right\}}
\newcommand{\sset}[2]{\left\{~#1  \left|
      \begin{array}{l}#2\end{array}
    \right.     \right\}}
\newcommand\tuple[1]{\langle #1 \rangle}

     {\end{tabular}\end{tt}\end{small}}

\newcommand{\false}{\mathit{false}}
\renewcommand{\true}{\mathit{true}}

\newcommand{\ignore}[1]{}

\newcommand{\bee}{\textsf{BEE}}

\begin{document}
\title{Compiling Finite Domain
  Constraints to\\ SAT with \bee: the Director's Cut}

\author{Michael Codish\and Yoav Fekete\and Amit Metodi}

\institute{ Department of Computer Science, Ben-Gurion University,
  Israel }

\maketitle
\label{firstpage}

\begin{abstract}
  \bee\ is a compiler which facilitates solving finite domain
  constraints by encoding them to CNF and applying an underlying SAT
  solver. In \bee\ constraints are modeled as Boolean functions which
  propagate information about equalities between Boolean literals.
  This information is then applied to simplify the CNF encoding of the
  constraints. We term this process \emph{equi-propagation}.
  A key factor is that considering only a small fragment of a
  constraint model at one time enables to apply stronger, and even
  complete reasoning to detect equivalent literals in that
  fragment. Once detected, equivalences propagate to simplify the
  entire constraint model and facilitate further reasoning on other
  fragments.
  \bee\ is described in several recent papers: \cite{Metodi2011},
  \cite{bee2012} and \cite{jair2013}.
  In this paper, after a quick review of \bee, we elaborate on two
  undocumented details of the implementation: the hybrid encoding of
  cardinality constraints and complete equi-propagation. We then
  describe on-going work aimed to extend \bee\ to consider binary
  representation of numbers.
\end{abstract}

\section{Introduction}

\bee\ (Ben-Gurion Equi-propagation Encoder) is a tool which applies to
encode finite domain constraint models to CNF. \bee\ was first
introduced in~\cite{bee2012} and is further described in
\cite{jair2013}.  During the encoding process, \bee\ performs
optimizations based on equi-propagation~\cite{Metodi2011} and partial
evaluation to improve the quality of the target CNF.
\bee\ is implemented in (SWI) Prolog and can be applied in conjunction
with any SAT solver. It can be downloaded from~\cite{bee2012web} where
one can also find examples of its use. This version of \bee\ is
configured to apply the CryptoMiniSAT solver \cite{Crypto} through a
Prolog interface \cite{satPearl}. CryptoMiniSAT offers direct support
for \texttt{xor} clauses, and \bee\ can be configured to take
advantage of this feature.

A main design choice of \bee\ is that integer variables are
represented in the unary order-encoding (see,
e.g.~\cite{baker,BailleuxB03}) which has many nice properties when
applied to small finite domains.
In the \emph{order-encoding}, an integer variable $X$ in the domain
$[0,\ldots,n]$ is represented by a bit vector $X=[x_1,\ldots,x_n]$.
Each bit $x_i$ is interpreted as $X\geq i$ implying that $X$
is a monotonic non-increasing Boolean sequence.
For example, the value 3 in the interval $[0,5]$ is represented in
5 bits as $[1,1,1,0,0]$.

It is well-known that the order-encoding facilitates the propagation
of bounds. Consider an integer variable $X=[x_1,\ldots,x_n]$ with
values in the interval $[0,n]$.  To restrict $X$ \pagebreak to take values in the
range $[a,b]$ (for $1\leq a\leq b\leq n$), it is sufficient to assign
$x_{a}=1$ and $x_{b+1}=0$ (if $b<n$). The variables $x_{a'}$ and
$x_{b'}$ for $1\leq a'< a$ and $b<b'\leq n$ are then determined $\true$
and $\false$, respectively, by \emph{unit propagation}.  For example, 
given $X=[x_1,\ldots,x_9]$, assigning $x_3=1$ and $x_6=0$ propagates
to give $X=[1,1,1,x_4,x_5,0,0,0,0]$, signifying that
$dom(X)=\{3,4,5\}$.
We observe an additional property of the order-encoding for
$X=[x_1,\ldots,x_n]$: its ability to specify that a variable cannot
take a specific value $0\leq v\leq n$ in its domain by equating two
variables: $x_{v}=x_{v+1}$.
This indicates that the order-encoding is well-suited not only to
propagate lower and upper bounds, but also to represent integer
variables with an arbitrary, finite set, domain.
For example, given $X=[x_1,\ldots,x_9]$, equating $x_2=x_3$ imposes
that $X\neq 2$. Likewise $x_5=x_6$ and $x_7=x_8$ impose that $X\neq 5$
and $X\neq 7$. Applying these equalities to $X$ gives,
$X=[x_1,\underline{x_2,x_2},x_4,\underline{x_5,x_5},\underline{x_7,x_7},x_9]$
(note the repeated literals),
signifying that $dom(X)=\{0,1,3,4,6,8,9\}$.

The order-encoding has many additional nice features that can be
exploited to simplify constraints and their encodings to CNF. To
illustrate one, consider a constraint of the form $\mathtt{A+B=5}$
where \texttt{A} and \texttt{B} are integer values in the range
between 0 and 5 represented in the order-encoding. At the bit level
(in the order encoding) we have: $\mathtt{A=[a_1,\ldots,a_5]}$ and
$\mathtt{B=[b_1,\ldots,b_5]}$.  The constraint is satisfied precisely
when $\mathtt{B=[\neg a_5,\ldots,\neg a_1]}$. Equi-propagation derives
the equations $E=\{b_1=\neg a_5,\ldots, b_5=\neg a_1\}$ and instead of
encoding the constraint to CNF, we apply the substitution indicated by
$E$, and remove the constraint which is a tautology given $E$.

\section{Compiling Constraints with \bee}

\bee\ is a constraint modeling language similar to the subset of
FlatZinc~\cite{miniZinc2007} relevant for finite domain constraint
problems. The full language is presented in Table \ref{tab:beeSyntax}.
Boolean constants ``$\true$'' and ``$\false$'' are viewed as (integer)
values ``1'' and ``0''.  Constraints are represented as (a list of)
Prolog terms. Boolean and integer variables are represented as Prolog
variables and may be instantiated when simplifying constraints.
In Table \ref{tab:beeSyntax}, $\mathtt{X}$ and $\mathtt{Xs}$ (possibly
with subscripts) denote a Boolean literal and a vector of literals,
$\mathtt{I}$ (possibly with subscript) denotes an integer variable,
and $\mathtt{c}$ (possibly with subscript) denotes an integer
constant.  On the right column of the table are brief explanations
regarding the constraints. The table introduces 26 constraint
templates.

Constraints (1-2) are about variable declarations: Booleans and
integers. Constraint (3) expresses a Boolean as an integer value.
Constraints (4-8) are about Boolean (and reified Boolean)
statements. The special cases of Constraint (5) for
$\mathtt{bool\_array\_or([X_1,\ldots,X_n])}$ and
$\mathtt{bool\_array\_xor([X_1,\ldots,X_n])}$ facilitate the
specification of clauses and of \texttt{xor} clauses (supported
directly in the CryptoMiniSAT solver~\cite{Crypto}).
Constraint (8) specifies that sorting a bit pair $\mathtt{[X_1,X_2]}$
(decreasing order) results in the pair $\mathtt{[X_3,X_4]}$. This is a
basic building block for the construction of sorting networks
\cite{Batcher68} used to encode cardinality (linear Boolean)
constraints during compilation as described in~\cite{AsinNOR11}
and~\cite{DBLP:conf/lpar/CodishZ10}.
Constraints (9-14) are about integer relations and operations.
Constraints (15-20) are about linear (Boolean, Pseudo Boolean, and
integer) operations.
Constraints (21-26) are about lexical orderings of Boolean and integer
arrays.

%


\begin{table}[t]\footnotesize
  \centering
\begin{tabular}{rlll}
\hline\hline
\multicolumn{4}{l}{\bf\small Declaring Variables}\\
\hline
(1)&$\mathtt{new\_bool(X)}$ & &declare Boolean \texttt{X}
\\
(2)&$\mathtt{new\_int(I,c_1,c_2)}$ & & declare integer \texttt{I}, $\mathtt{c_1\leq I\leq c_2}$
\\
(3)&    $\mathtt{bool2int(X,I)}$ &
          &
          $\mathtt{(X  \Leftrightarrow I = 1) \wedge (\neg X \Leftrightarrow I = 0)}$\\
\hline

\multicolumn{4}{l}{\bf\small Boolean (reified) Statements~ 
       \hfill $\mathtt{op\in\{or, and, xor, iff\}}$}\\
\hline
(4)&    $\mathtt{bool\_eq(X_1,X_2)}$ ~or~ $\mathtt{bool\_eq(X_1,-X_2)}$&
          $\mathtt{}$&
          $\mathtt{X_1 = X_2}$ ~or~ $\mathtt{X_1 = -X_2}$\\
(5)&    $\mathtt{bool\_array\_op([X_1,\ldots,X_n])}$ &
          $\mathtt{}$&
          $\mathtt{X_1 ~op~ X_2 \cdots op~ X_n}$\\
(6)&    $\mathtt{bool\_array\_op\_reif([X_1,\ldots,X_n],~X)}$ &
          $\mathtt{}$&
          $\mathtt{X_1 ~op~ X_2 \cdots op~ X_n\Leftrightarrow X}$\\
(7)&    $\mathtt{bool\_op\_reif(X_1,X_2,~X)}$ &
          $\mathtt{}$&
          $\mathtt{X_1 ~op~ X_2\Leftrightarrow X}$\\
(8)&$\mathtt{comparator(X_1,X_2,X_3,X_4)}$ &$\mathtt{}$&
          $\mathtt{sort([X_1,X_2])=[X_3,X_4]}$\\
\hline

\multicolumn{4}{l}{\bf\small Integer relations (reified)
                    \hfill $\mathtt{rel\in\{leq, geq, eq, lt, gt, neq\}}$}\\
\multicolumn{4}{l}{\bf\small \& arithmetic \hfill
            ~$\mathtt{op\in\{plus, times, div, mod, max, min\}}$, 
             $\mathtt{op'\in\{plus, times, max, min\}}$  }\\
    \hline
(9)&    $\mathtt{int\_rel(I_1,I_2)}$ &
          $\mathtt{}$&
          $\mathtt{I_1 ~rel~ I_2}$\\
(10)&    $\mathtt{int\_rel\_reif(I_1,I_2,~X)}$ &
          $\mathtt{}$&
          $\mathtt{I_1 ~rel~ I_2 \Leftrightarrow X}$\\
(11)&    $\mathtt{int\_array\_allDiff([I_1,\ldots,I_n])}$ &
          &
          $\mathtt{\bigwedge_{i<j}I_i \neq I_j}$\\

(12)&$\mathtt{int\_abs(I_1,~I)}$ &
      $\mathtt{}$&
      $\mathtt{|I_1| = I}$\\
(13)&$\mathtt{int\_op(I_1,I_2,~I)}$ &
      $\mathtt{}$&
      $\mathtt{I_1 ~op~ I_2 = I}$\\
(14)&$\mathtt{int\_array\_op'([I_1,\ldots,I_n],~I)}$ &
      $\mathtt{}$&
      $\mathtt{I_1 ~op'\cdots op'~ I_n = I}$\\
\hline

\multicolumn{4}{l}{\bf\small  Linear Constraints~
            \hfill $\mathtt{rel{\in}\{leq, geq, eq, lt, gt\}}$}\\
    \hline
(15)&    $\mathtt{bool\_array\_sum\_rel([X_1,\ldots,X_n],~I)}$ &
          $\mathtt{}$&
          $\mathtt{(\Sigma ~X_i)~ rel~ I}$\\
(16)&$\mathtt{bool\_array\_pb\_rel([c_1,\ldots,c_n],[X_1,\ldots,X_n],~I)}$ &
      $\mathtt{}$&
          $\mathtt{(\Sigma ~c_i*X_i)~ rel~ I}$\\
(17)&$\mathtt{bool\_array\_sum\_modK([X_1,\ldots,X_n],c,~I)}$ &
      $\mathtt{}$&
          $\mathtt{((\Sigma ~X_i)~ mod~ c) =~ I}$\\

(18)&$\mathtt{int\_array\_sum\_rel([I_1,\ldots,I_n],~I)}$ &
      $\mathtt{}$&
          $\mathtt{(\Sigma ~I_i)~ rel~ I}$\\
(19)&$\mathtt{int\_array\_lin\_rel([c_1,\ldots,c_n],[I_1,\ldots,I_n],~I)}$ &
      $\mathtt{}$&
          $\mathtt{(\Sigma ~c_i*I_i)~ rel~ I}$\\

(20)&$\mathtt{int\_array\_sum\_modK([I_1,\ldots,I_n],c,~I)}$ &
      $\mathtt{}$&
          $\mathtt{((\Sigma ~I_i)~ mod~ c) =~ I}$\\
\hline

\multicolumn{4}{l}{\bf\small  Lexical Order}\\
    \hline

(21)&     $\mathtt{bool\_arrays\_lex(Xs_1,Xs_2)}$ &$\mathtt{}$&
             $\mathtt{Xs_1} \preceq \mathtt{Xs_2}$ (lex order)\\
(22)&     $\mathtt{bool\_arrays\_lexLt(Xs_1,Xs_2)}$ &$\mathtt{}$&
             $\mathtt{Xs_1}\prec\mathtt{Xs_2}$ (lex order)\\
(23)&     $\mathtt{bool\_arrays\_lex\_reif(Xs_1,Xs_2,X)}$ &$\mathtt{}$&
             $\mathtt{X \Leftrightarrow}\mathtt{Xs_1}\preceq \mathtt{Xs_2}$\\
(24)&     $\mathtt{bool\_arrays\_lexLt\_reif(Xs_1,Xs_2,X)}$ &$\mathtt{}$&
            $\mathtt{X \Leftrightarrow}\mathtt{Xs_1}\prec\mathtt{Xs_2}$ \\

(25)&     $\mathtt{int\_arrays\_lex(Is_1,Is_2)}$ &$\mathtt{}$&
             $\mathtt{Is_1}\preceq\mathtt{Is_2}$ (lex order)\\
(26)&     $\mathtt{int\_arrays\_lexLt(Is_1,Is_2)}$ &$\mathtt{}$&
             $\mathtt{Is_1}\prec\mathtt{Is_2}$ (lex order)\\

\hline\hline
\end{tabular}
  \caption{Syntax of \bee\ Constraints. }
  \label{tab:beeSyntax}
\end{table}

The compilation of a constraint model to a CNF
using \bee\ goes through three phases:
%
\textbf{(1)}~Unary bit-blasting: integer variables (and constants) are
  represented as bit vectors in the order-encoding.
\textbf{(2)}~Constraint simplification: three types of actions are applied:
  equi-propagation, partial evaluation, and decomposition of
  constraints.  Simplification is applied repeatedly until no rule is
  applicable.
\textbf{(3)}~CNF encoding: the best suited encoding technique is applied to
  the simplified constraints.

Bit-blasting is implemented through Prolog unification. Each
declaration of the form $\mathtt{new\_int(I,c_1,c_2)}$ triggers a
unification $\mathtt{I=[1,\dots,1,X_{c_1+1},\ldots,X_{c_2}]}$. To ease
presentation we assume that integer variables are represented in a
positive interval starting from~$0$ but there is no such limitation in
practice (\bee\ supports also negative integers).
\bee\ applies ad-hoc equi-propagators.  Each constraint is associated
with a set of ad-hoc rules.  The novelty is that the approach is not
based on CNF, as in previous works (for example
\cite{chu-min2003}, \cite{sateliteEenB05}, 
\cite{HeuleJarvisaloBiere2011}, and \cite{Manthey2012}),
but rather driven by the bit
blasted constraints that are to be encoded to CNF.
For example, Figure~\ref{fig:rules2} illustrates the rules for the
\plus\ constraint.
\begin{wrapfigure}[12]{r}{72mm}
\small
\vspace{-5mm}
\begin{tabular}{|c|c|}
\hline
  \multicolumn{2}{|c|}{\small $c=\plus(X,Y,Z)$ where $X=\tuple{x_1,\ldots,x_n}$,}\\
  \multicolumn{2}{|c|}{\small $Y=\tuple{y_1,\ldots,y_m}$, and
                                $Z=\tuple{z_1,\ldots,z_{n+m}}$}\\
\hline
  {if } & {then propagate} \\
\hline
\hline
$X\geq i$, $Y\geq j$ & $Z\geq i+j$\\
\hline
$X<i$, $Y<j$ & $Z<i+j-1$\\
\hline
$Z \geq k$, $X < i$ & $Y \geq k-i$ \\
\hline
$Z < k$, $X \geq i$ & $Y < k-i$ \\
\hline
$X=i$ & $z_{i+1}=y_1,\ldots,z_{i+m}=y_m$ \\
\hline
$Z=k$ &  $x_1=\neg y_k,\ldots,x_k=\neg y_1$ \\
\hline 
\end{tabular}
\caption{Ad-hoc rules for  $\plus$}
\label{fig:rules2}
\end{wrapfigure}
For an integer $X=\tuple{x_1,\ldots,x_n}$, we
write: $X\geq i$ to denote the equation $x_i=1$, $X<i$ to denote the
equation $x_i=0$, $X\neq i$ to denote the equation $x_i=x_{i+1}$, and
$X=i$ to denote the pair of equations $x_i=1,x_{i+1}=0$.
We view $X=\tuple{x_1,\ldots,x_n}$ as if padded with sentinel cells
such that all cells ``to the left of'' $x_1$ take value 1 and all
cells ``to the right of'' $x_n$ take the value 0. This facilitates the
specification of the ``end cases'' in the formalism.
The first four rules of Figure~\ref{fig:rules2} capture the
standard propagation behavior for interval arithmetic. The last two
rules apply when one of the integers in the relation is a
constant. There are symmetric cases when replacing the role of $X$ and
$Y$.

When an equality of the form $X=L$ (between a variable and a literal
or a constant) is detected, then equi-propagation is implemented by
unifying $X$ and $L$ and applies to all occurrences of $X$ thus
propagating to other constraints involving $X$.

Decomposition is about replacing complex constraints (for example
about arrays) with simpler constraints (for example about array
elements). Consider, for instance, the constraint
$\mathtt{int\_array\_plus(As,Sum)}$. It is decomposed to a list of
$\mathtt{int\_plus}$ constraints applying a straightforward divide and
conquer recursive definition. At the base case, if \texttt{As=[A]}
then the constraint is replaced by a constraint of the form
\texttt{int\_eq(A,Sum)} which equates the bits of $\mathtt{A}$ and
$\mathtt{Sum}$, or if $\mathtt{As=[A_1,A_2]}$ then it is replaced by
$\mathtt{int\_plus(A_1,A_2,Sum)}$.
In the general case \texttt{As} is split into two halves, then
constraints are generated to sum these halves, and then an additional
$\mathtt{int\_plus}$ constraint is introduced to sum the two sums.

CNF encoding is the last phase in the compilation of a constraint
model. Each of the remaining simplified (bit-blasted) constraints is
encoded directly to a CNF. These encodings are standard and similar to
those applied in various tools such as Sugar~\cite{sugar2009}.

\section{Cardinality Constraints in \bee}

Cardinality Constraints take the form $\sum \{x_1,\ldots,x_n\} \leq k$
where the $x_i$ are Boolean literals, $k$ is a constant, and the
relation $\leq$ might be any of $\{=,<,>,\leq,\geq\}$.  There is a
wide body of research on the encoding of cardinality to CNF. We focus
on those using sorting networks. For example, the presentations in
\cite{EenS06}, \cite{CodishFFS11}, and \cite{AsinNOR09,AsinNOR11}
describe the use of odd-even sorting networks to encode pseudo Boolean
and cardinality constraints to Boolean formula.
We observe that for applications of this type, it suffices to apply
``selection networks''~\cite{Knuth73} rather than sorting
networks. Selection networks apply to select the $k$ largest
elements from $n$ inputs.  In~\cite{Knuth73}, Knuth shows a simple
construction of a selection network with $O(n \log^2 k)$ size whereas,
the corresponding sorting network is of size $O(n\log^2n)$.
Totalizers~\cite{BailleuxB03} are similar to sorting networks except
that the merger for two sorted sequences  involves a direct
encoding with $O(n^2)$ clauses instead of  $O(n\log
n)$ clauses. Totalizers have been shown to give better encodings when
cardinality constraints are not excessively large.
\bee\  enables the user to select encodings based on sorting networks,
totalizers or a hybrid approach which is further detailed below.

Consider the constraint $\mathtt{bool\_array\_sum\_eq(As,Y)}$ in a
context where $\mathtt{As}$ is a list of $n$ Boolean literals and
integer variable $\mathtt{Y}$ defined as
$\mathtt{new\_int(Y,0,n)}$. \bee\ applies a divide and conquer
strategy. If $n=1$, the constraint is trivial and satisfied by
unifying $\mathtt{Y=As}$. If $n=2$ and $\mathtt{As=[A_1,A_2]}$ then
$\mathtt{Ys=[Y_1,Y_2]}$ and the constraint is decomposed to
$\mathtt{comparator(A_1,A_2,Y_1,Y_2)}$. In the general case, where
$n>2$, the constraint is decomposed as follows where $\mathtt{As_1}$
and $\mathtt{As_2}$ are a partitioning of $\mathtt{As}$ such that
$\mathtt{|As_1|=n_1}$, $\mathtt{|As_2|=n_2}$, and
$\mathtt{|n_1-n_2|\leq 1}$:
\[\small
\fbox{$\mathtt{bool\_array\_sum\_eq(As,Y)}$}
\xRightarrow[]{\mathtt{decomp.}}
\fbox{$
  \begin{array}{ll}
    \mathtt{new\_int(T_1,0,n_1)},\quad & \mathtt{bool\_array\_sum\_eq(As_1,T_1)},\\
    \mathtt{new\_int(T_2,0,n_2)},\quad & \mathtt{bool\_array\_sum\_eq(As_2,T_2)},\\
    \mathtt{int\_plus(T_1,T_2,Y)}  
  \end{array}$}
\] 
This decomposition process continues as long as there remain
$\mathtt{bool\_array\_sum\_eq}$ and when it terminates the model contains
only \texttt{comparator} and \texttt{int\_plus} constraints.
The interesting discussion is with regards to the $\mathtt{int\_plus}$
constraints where \bee\ offers two options and depending on this
choice the original \texttt{bool\_array\_sum\_eq} constraint then
takes the form either of a sorting network~\cite{Batcher68} or of a
totalizer~\cite{BailleuxB03}.
So, consider a  constraint $\mathtt{int\_plus(A,B,C)}$ where
$\mathtt{A=[A_1,\ldots,A_m]}$, $\mathtt{B=[B_1,\ldots,B_p]}$ and
$\mathtt{C=[C_1,\ldots,C_{m+p}]}$ represent integer variables in the
order encoding.
A unary adder leads to a direct encoding of the sum of two unary
numbers. It involves $O(n^2)$  clauses where $n$ is the size of
the inputs and as a circuit it has ``depth'' 1. The encoding introduces
the following clauses where $(1\leq i\leq m)$ and $(1\leq j\leq p)$:
\[\small
\begin{array}{llll}
  \bullet & \bigwedge_i \mathtt{(A_i \rightarrow C_i)} &
  \bullet & \bigwedge_i \mathtt{(\neg A_i \rightarrow \neg C_{p+i})}\\
  \bullet & \bigwedge_j \mathtt{(B_j \rightarrow C_j)} &
  \bullet & \bigwedge_j \mathtt{(\neg B_j \rightarrow \neg C_{m+j})}\\
  \bullet & \bigwedge_{i,j} \mathtt{(A_i \wedge B_j \rightarrow C_{i+j})} \qquad& 
  \bullet & \bigwedge_{i,j} \mathtt{(\neg A_i \wedge \neg B_j \rightarrow \neg C_{i+j-1})}\\
\end{array}
\]
An alternative encoding for $\mathtt{int\_plus(A,B,C)}$ is obtained by
means of a recursive decomposition based on the so called odd-even
merger from Batcher's construction~\cite{Batcher68}. It leads to an
encoding with $O(n\log n)$ clauses where $n$ is the size of the inputs
and as a circuit it has ``depth'' $\log n$. The decomposition is as
follows (ignoring the base cases) where $\mathtt{A_{o}}$,
$\mathtt{A_{e}}$, $\mathtt{B_{o}}$ and $\mathtt{B_{e}}$ are partitions
of $\mathtt{A}$ and $\mathtt{B}$ to their odd and even positioned
elements, $\mathtt{C_o}$, $\mathtt{C_e}$ are new unary variables
defined with the appropriate domains, and where
$\mathtt{combine(C_o,C_e,C)}$ signifies a set of comparator
constraints and is defined as $\bigwedge_{i}
\mathtt{comparator(C_{o_{i+1}},C_{e_i},C_{2i},C_{2i+1})}$:
\[\small
\fbox{$\mathtt{int\_plus(A,B,C)}$}
\xRightarrow[]{\mathtt{decompose}}
\fbox{$
  \begin{array}{l}
     \mathtt{int\_plus(A_o,B_o,C_o)},\\
     \mathtt{int\_plus(A_e,B_e,C_e)},\\
     \mathtt{combine(C_o,C_e,C)}
  \end{array}$}
\] 
%
%
In addition to the encodings based on unary adders (direct) and
mergers (recursive decomposition), \bee\ offers a combination of the
two which we call ``hybrid''.  The intuition is simple: in the hybrid
approach we perform recursive decomposition as for odd-even mergers,
but only so long as the resulting CNF is predetermined to be
smaller than the corresponding unary adder. So, it is just like a
merger except that the base case is a unary adder.
Before each decomposition of $\mathtt{int\_plus}$, \bee\ evaluates the
benefit (in terms of CNF size) of decomposing the constraint as a
merger and takes the smaller of the two.

\begin{figure}[t]
\begin{minipage}{0.47\linewidth}
  \includegraphics[width=1.0\linewidth,keepaspectratio]{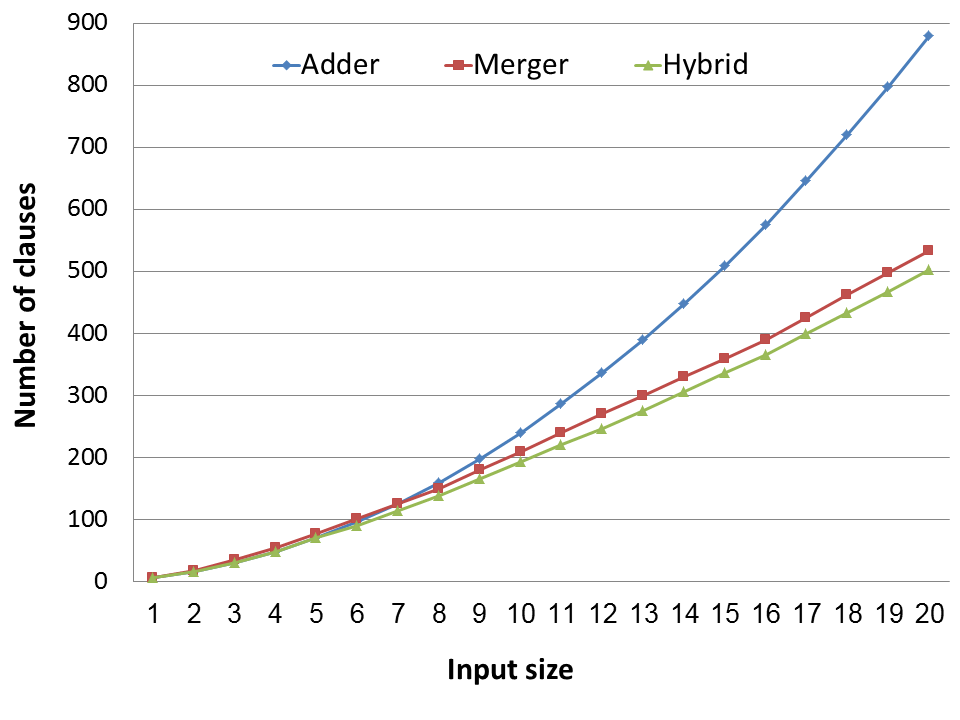}
\end{minipage}
\qquad
\begin{minipage}{0.47\linewidth}
\includegraphics[width=1.0\linewidth,keepaspectratio]{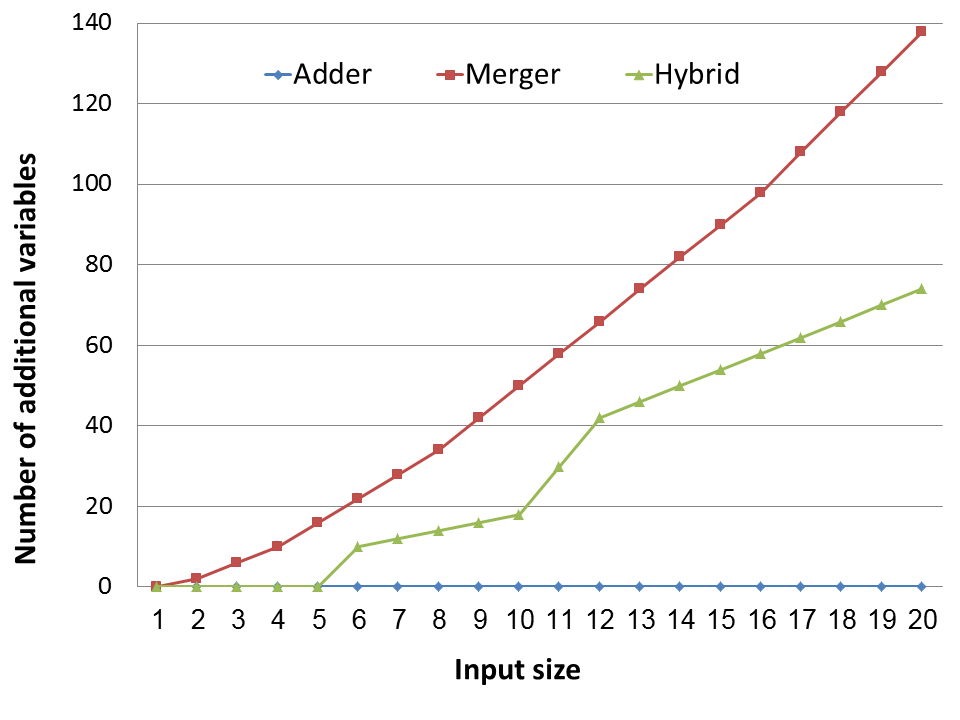}
\end{minipage}
\caption {Relative size of CNF encodings for cardinality: adders,
  hybrid \& mergers. On the left number of clauses, and on the right
  number of added variables.}
 \label{fig:size}
\end{figure}


Figure~\ref{fig:size} depicts the size of CNF encodings for the
constraint $\mathtt{int\_plus(A,B,C)}$ where
$|\mathtt{A}|=|\mathtt{B}|=n$. The left graph illustrates the number
of clauses in the three encodings. The unary adder has fewest number
of clauses for inputs of size 7 or less. The hybrid encoding is always
just slightly smaller than the merger. Each time a merger is
decomposed to an adder it is just about of the same number of clauses.
In contrast, in the right graph we see that the encoding never
introduces fresh variables, and as the size of the input increases so
does the benefit of the hybrid approach in number of added variables.



Now let us consider the constraint
$\mathtt{bool\_array\_sum\_leq(As,k)}$ where $\mathtt{As}$ is a list
of $n$ Boolean literals and $k$ is a constant.  Assume as before that
$\mathtt{As_1}$ and $\mathtt{As_2}$ are a partitioning of
$\mathtt{As}$ such that $\mathtt{|As_1|=n_1}$, $\mathtt{|As_2|=n_2}$,
and $\mathtt{|n_1-n_2|\leq 1}$.
A naive decomposition might proceed as follows:\pagebreak
\[\small
\begin{array}{r}
    \fbox{$\mathtt{bool\_array\_sum\_leq(As,k)}$}
    \xRightarrow[]{\mathtt{decompose}}
    \fbox{$
      \begin{array}{l}
        \mathtt{new\_int(Y,0,n)},\\
        \mathtt{bool\_array\_sum\_eq(As,Y)},\\
        \mathtt{int\_leq(Y,k)}
      \end{array}$}
\\[8mm]
    \xRightarrow[]{\mathtt{decompose}}
    \fbox{$\begin{array}{ll}
    \mathtt{new\_int(Y,0,n)},\\
    \mathtt{new\_int(T_1,0,n_1)},~ & \mathtt{bool\_array\_sum\_eq(As_1,T_1)},\\
    \mathtt{new\_int(T_2,0,n_2)},~ & \mathtt{bool\_array\_sum\_eq(As_2,T_2)},\\
    \mathtt{int\_plus(T_1,T_2,Y)},&
    \mathtt{int\_leq(Y,k)}
  \end{array}$}
\end{array}
\] 
But we can do better.
In \bee\ we decompose $\mathtt{bool\_array\_sum\_leq(As,k)}$ as
follows:
\[\small \fbox{$\mathtt{b.\_a.\_sum\_leq(As,k)}$}
    \xRightarrow[]{\mathtt{decomp.}}
    \fbox{$
  \begin{array}{ll}
    \mathtt{new\_int(T_3,0,k)},\\
    \mathtt{new\_int(T_1,0,min(n_1,k))}, & 
                   \mathtt{bool\_array\_sum\_eq(As_1,T_1)},\\
    \mathtt{new\_int(T_2,0,min(n_2,k))}, & 
                   \mathtt{bool\_array\_sum\_eq(As_2,T_2)},\\
     \mathtt{int\_plus(T_3,T_2,k)}, & \mathtt{int\_leq(T_1,T_3)}\\
    
  \end{array}$}
\]
This is correct because the constraint
$\mathtt{int\_plus(T_3,T_2,k)}$ defines $T_3 = k-T_2$ and so we have
\[(T_1 + T_2 \leq k)    \leftrightarrow     
  (T_1 \leq T_3) \wedge  (T_2+T_3=k)
\]
This encoding is preferable  because the
$\mathtt{int\_plus(T_3,T_2,k)}$ constraint is encoded with 0 clauses
(due to equi-propagation) and the $\mathtt{int\_leq(T_1,T_3)}$
constraint in $O(k)$ clauses. Whereas in the naive version the
$\mathtt{int\_plus(T_1,T_2,Y)}$ is encoded in $O(n \log(n))$ or
$O(n^2)$ (sorting network or direct) and the $\mathtt{int\_leq(Y,k)}$
is encoded with 0 clauses.
 

\section{Complete Equi-Propagation in \bee}

Equi-propagation is about inferring Boolean Equalities, $x=\ell$,
implied from a given CNF formula $\varphi$ where $x$ is a Boolean
variable and $\ell$ a Boolean constant or literal. 
Complete equi-propagation (CEP) is about inferring all such
equalities.  Equi-propagation in \bee\ is based on ad-hoc rules and
thus incomplete. However, \bee\ allows the user to specify, for given
sets of constraints in a model, that CEP is to be applied (instead of
ad-hoc equi-propagation).
CEP generalizes the notion of a backbone \cite{Schneider96}. The
backbone of a CNF, $\varphi$, is the set of literals that are true in
all models of $\varphi$, thus corresponding to the subset of
equations, $x=\ell'$ obtained from CEP where $\ell'$ is a Boolean
constant. Backbones prove useful in applications of SAT such as model
enumeration, minimal model computation, prime implicant computation,
and also in applications which involve optimization (see for example,
\cite{Marques-SilvaJL10}).  Assigning values to backbone variables
reduces the size of the search space while maintaining the meaning of
the original formula. In exactly the same way, CEP identifies
additional variables that can be removed from a formula, to reduce the
search space, by equating pairs of literals, as in $x=y$ or $x=-y$.

Backbones are often computed by iterating with a SAT solver.
In~\cite{Marques-SilvaJL10}, the authors describe and evaluate several
such algorithms and present an improved algorithm. This algorithm
involves\footnote{See Proposition~6 in
  \url{http://sat.inesc-id.pt/~mikolas/bb-aicom-preprint.pdf}.}
exactly one unsatisfiable call to the sat solver and at most $n-b$
satisfiable calls, where $n$ is the number of variables in $\varphi$
and $b$ the size of its backbone.

It is straightforward to apply an algorithm that computes the backbone
of a CNF, $\varphi$, to perform CEP (to detect also equations between
literals). Enumerating the variables of $\varphi$ as
$\set{x_1,\ldots,x_n}$. One simply defines
\begin{equation}
  \label{eq:cep}
  \varphi' =
         \varphi\land
         \sset{e_{ij}\leftrightarrow (x_i\leftrightarrow x_j)}
              {0\leq i<j\leq n}
\end{equation}
introducing $\theta(n^2)$ fresh variables $e_{ij}$. If $e_{ij}$ is in
the backbone of $\varphi'$ then $x_i=x_j$ is implied by $\varphi$, and if
$\neg e_{ij}$ is in the backbone then $x_i= \neg x_j$ is implied.
A major obstacle is that computing the backbone of $\varphi$ is at
least as hard as testing for the satisfiability of $\varphi$
itself. Hence, for \bee, the importance of the assumption that
$\varphi$ is only a small fragment of the CNF of interest. Another
obstacle is that the application of CEP for $\varphi$ with $n$
variables involves computing the backbone for $\varphi'$ which has
$\theta(n^2)$ variables.

The CEP algorithm applied in \bee\ is basically the same as that
proposed for computing backbones in~\cite{Marques-SilvaJL10} extending
$\varphi$ to $\varphi'$ as prescribed by Equation (\ref{eq:cep}). We
prove that iterated SAT solving for CEP using $\varphi'$ involves at
most $n+1$ satisfiable SAT tests, and exactly one unsatisfiable test,
in spite of the fact that $\varphi'$ involves a quadratic number of
fresh variables.

We first describe the algorithm applied to compute the backbone of a
given formula $\varphi$, which we assume is satisfiable.
The algorithm maintains a table indicating for each variable $x$ in
$\varphi$ for which values of $x$, $\varphi$ can be satisfied:
$\true$, $\false$, or both. The algorithm is initialized by calling
the SAT solver with $\varphi_0 = \varphi$ and initializing the table
with the information relevant to each variable: if the solution for
$\varphi_0$ assigns a value to $x$ then that value is tabled for
$x$. If it assigns no value to $x$ then both values are tabled for
$x$.

The algorithm iterates incrementally. For each step $i>0$ we add a
single clause $C_i$ (detailed below) and reinvoke the same solver
instance, maintaining the learned data of the solver. This process
terminates with a single unsatisfiable invocation. In words: the
clause $C_i$ can be seen as asking the solver if it is possible to
flip the value for any of the variables for which we have so far seen
only a single value.
More formally, at each step of the algorithm, $C_i$ is defined as
follows: for each variable $x$, if the table indicates a single value
$v$ for $x$ then $C_i$ includes $\neg v$. Otherwise, if the table
indicates two values for $x$ then there is no corresponding literal in
$C_i$. The SAT solver is then called with $\varphi_i =
\varphi_{i-1}\wedge C_i$. If this call is satisfiable then the table
is updated to record new values for variables (there must be at least
one new value in the table) and we iterate. Otherwise, the algorithm
terminates and the variables remaining with single entries in the
table are the backbone of $\varphi$.

\begin{figure}
\centering
\subfloat[][Demo of backbone algorithm (Example~\ref{example:cep1})]{
{\small
  $\begin{array}[t]{@{\extracolsep{-.6mm}}l|lllll|l}
    &   x_1 & x_2 & x_3  & x_4 & x_5 & \\
    \hline
 \theta_1 & 1 & 1 & 0 & 0 & 1 &
            \varphi_0 = \varphi\\
 \theta_2   & 1 & 0 & 0 & 1 & 0 &
            \varphi_1 = \varphi_0 \wedge \neg\theta_1 \\
 \theta_3 & \multicolumn{5}{c|}{unsat} &
            \varphi_2 = \varphi_1 \wedge 
            (\neg x_1 \vee   x_3)\\
\end{array}$
}
\label{fig:subfig1}}
\qquad
\subfloat[][Demo of proof that CEP is linear (Example~\ref{example:cep3})]{
{\small
  $\begin{array}[t]{@{\extracolsep{-.6mm}}l|lllll|l}
    &   x_1 & x_2 & x_3  & x_4 & x_5 & \\
    \hline
 \theta_1 & 1 & 1 & 0 & 0 & 1 &
            \set{x_1,x_2,x_3,x_4,x_5,1}\\
 \theta_2   & 1 & 0 & 0 & 1 & 0 &
            \set{x_1,x_3,1}, \set{x_2,x_4,x_5}\\
 \theta_3   & 1 & 0 & 0 & 0 & 1 & 
            \set{x_1,x_3,1}, \set{x_2},\set{x_4,x_5}\\
 \theta_4 & \multicolumn{5}{c|}{unsat} &
            \\
\end{array}$
}
\label{fig:subfig2}}

\subfloat[][Demo of the CEP algorithm (Example~\ref{example:cep2})]{
{\small
  $\begin{array}{@{\extracolsep{-.5mm}}c|ccccc|cccccccccc|l}
    &x_1&x_2&x_3&x_4&x_5&e_{12}&e_{13}&e_{14}&e_{15}&
                              e_{23}&e_{24}&e_{25}&
                              e_{34}&e_{35}&e_{45}&\\
    \hline
 \theta_1 & 1 & 1 & 0 & 0 & 1 &    1&0&0&1& 0&0&1& 1&0& 0 &  
            \varphi_0 = \varphi\\
 \theta_2 & 1 & 0 & 0 & 1 & 0 &    0&0&1&0& 1&0&1& 0&1& 0 &
            \varphi_1 = \varphi_0 \wedge \neg\theta_1 \\
 \theta_3 & 1 & 0 & 0 & 0 & 1 &    0&0&0&1& 1&1&0& 1&0& 0 &
           \varphi_2 = \varphi_1 \wedge 
          \left(
            \begin{array}{l}
              \neg x_1 \vee x_3 \vee e_{13} \vee \\
              \neg e_{24} \vee \neg e_{25} \vee e_{45} 
            \end{array}
          \right)\\
 \theta_4 & \multicolumn{5}{c|}{unsat} & \multicolumn{10}{c|}{}&
          \varphi_3=\varphi_2\wedge (\neg x_1 \vee x_3 \vee e_{13}\vee e_{45})\\
\end{array}$
}
\label{fig:subfig3}}

\caption{Demonstrating Examples~\ref{example:cep1}---\ref{example:cep3}}
\label{fig}
\end{figure}

\begin{example}\label{example:cep1}

  Figure~\ref{fig}\;\subref{fig:subfig1} where we assume given a
  formula, $\varphi$, which has models as indicated below illustrates
  the backbone algorithm.  The first two iterations of the algorithm
  provide the models, $\theta_1$ and $\theta_2$. The next iteration
  illustrates that $\varphi$ has no model which satisfies $\varphi$
  and flips the values of $x_1$ (to false) or of $x_3$ (to true). We
  conclude that $x_1$ and $x_3$ are the backbone variables of
  $\varphi$.

 \end{example}

Now consider the case where in addition to the backbone we wish to
derive also equations between literals which hold in all models of
$\varphi$. The CEP algorithm for $\varphi$ is as follows: (1)~enhance
$\varphi$ to $\varphi'$ as specified in Equation~\ref{eq:cep}, and (2)
apply backbone computation to $\varphi'$. 
If $\varphi'\models e_{xy}$ then $\varphi\models {x=y}$ and if
$\varphi'\models \neg e_{xy}$ then $\varphi\models {x= \neg y}$.
As an optimization, it is possible to focus in the first two
iterations only on the variables of $\varphi$.
\begin{example}\label{example:cep2}

  Consider the same formula $\varphi$ as in
  Example~\ref{example:cep1}.  This time, in the third iteration we
  ask to either flip the value for one of $\{x_1,x_3\}$ or for one of
  $\{e_{13},e_{24},e_{25},e_{45}\}$ and there is such a model, $\theta_3$. This is
  illustrated as Figure~\ref{fig}\;\subref{fig:subfig3}
\end{example}

\begin{theorem}\label{thm:cep}
  Let $\varphi$ be a CNF, $X$ a set of $n$ variables, and
  $\Theta=\set{\theta_1,\ldots,\theta_m}$ the sequence of assignments
  encountered by the CEP algorithm for $\varphi$ and $X$. Then, $m\leq
  n+1$.
\end{theorem}

Before presenting a proof of Theorem~\ref{thm:cep} we introduce some
terminology.
Assume a set of Boolean variables $X$ and a sequence
$\Theta=\set{\theta_1,\ldots,\theta_m}$ of models. Denote $\hat X=
X\cup\set{1}$ and let $x,y\in \hat X$.
If $\theta(x)=\theta(y)$ for all $\theta\in\Theta$ or if
$\theta(x)\neq\theta(y)$ for all $\theta\in\Theta$, then we say that
$\Theta$ \emph{determines} the equation $x=y$. Otherwise, we say that
$\Theta$ \emph{disqualifies} $x=y$, intuitively meaning that $\Theta$
disqualifies $x=y$ from being determined. More formally, $\Theta$
\emph{determines} $x=y$ if and only if $\Theta\models (x=y)$ or
$\Theta\models (x= \neg y)$, and otherwise $\Theta$
\emph{disqualifies} $x=y$.

The CEP algorithm for a formula $\varphi$ and set of $n$ variables $X$
applies so that each iteration results in a satisfying assignment for
$\varphi$ which disqualifies at least one additional equation between
elements of $\hat X$. Although there are a quadratic number of
equations to be considered, we prove that the CEP algorithm terminates
after at most $n+1$ iterations.

\begin{proof} (of Theorem~\ref{thm:cep})
  For each value $i\leq m$, $\Theta_i= \set{\theta_1, \ldots,
    \theta_i}$ induces a partitioning, $\Pi_i$ of $\hat X$ to disjoint
  and non-empty sets, defined such that for each $x,y\in \hat X$, $x$
  and $y$ are in the same partition $P\in \Pi_i$ if and only if
  $\Theta_i$ determines the equation $x=y$. So, if $x,y\in P\in\Pi_i$
  then the equation $x=y$ takes the same value in all assignments of
  $\Theta_i$. The partitioning is well defined because if in all
  assignments of $\Theta_i$ both $x=y$ takes the same value and $y=z$
  takes the same value, then clearly also $x=z$ takes the same value,
  implying that $x,y,z$ are in the same partition of $\Pi_i$.
  Finally, note that each iteration $1<i\leq m$ of the CEP algorithm
  disqualifies at least one equation $x=y$ that was determined by
  $\Theta_{i-1}$. This implies that at least one partition of
  $\Pi_{i-1}$ is split into two smaller (non-empty) partitions of
  $\Pi_{i}$. As there are a total of $n+1$ elements in $\hat X$, there
  can be at most $n+1$ iterations to the algorithm.
\end{proof}

\begin{example}\label{example:cep3}

  Consider the same formula $\varphi$ as in
  Examples~\ref{example:cep1}
  and~\ref{example:cep2}. Figure~\ref{fig}\;\subref{fig:subfig2}
  illustrates the run of the algorithm in terms of the partitioning
  $\Pi$ from the proof of Theorem~\ref{thm:cep}.

\end{example}

We illustrate the impact of CEP with an application where the goal is
to find the largest number of edges in a simple graph with $n$ nodes
such that any cycle (length) is larger than 4. The graph is
represented as a Boolean adjacency matrix $A$ and there are two types
of constraints: (1)~constraints about cycles in the graph:
$\forall_{i,j,k}.~ A[i,j]+A[j,k]+A[k,i] < 3$, and
$\forall_{i,j,k,l}.~A[i,j]+A[j,k]+A[k,l] + A[l,i] < 4$; and
(2)~constraints about symmetries: in addition to the obvious
$\forall_{1\leq i<j\leq n}.~(A[i,j]\equiv A[j,i] \mbox{~~ and~~ }
A[i,i]\equiv\false)$, we constrain the rows of the adjacency matrix to
be sorted lexicographically (justified in~\cite{cmps:ijcai13}), and we
impose lower and upper bounds on the degrees of the graph nodes as
described in~\cite{Garnick93}.

Table~\ref{table:cep} illustrates results, running \bee\ with and
without CEP. Here, we focus on finding a graph with the prescribed
number of graph nodes with the known maximal number of edges (all
instances are satisfiable), and CEP is applied to the set of clauses
derived from the symmetry constraints (2) detailed above.  The table
indicates the number of nodes, and for each CEP choice: the \bee\
compilation time, the number of clauses and variables, and the
subsequent sat solving time.  The table indicates that CEP increases
the compilation time (within reason), reduces the CNF size
(considerably), and (for the most) improves SAT solving
time.\footnote{\label{machine}Experiments are performed on a single
  core of an Intel(R) Core(TM) i5-2400 3.10GHz CPU with 4GB memory
  under Linux (Ubuntu lucid, kernel 2.6.32-24-generic).}

\begin{table}
  \centering\small
\begin{tabular}{|cr|rrrr|rrrr|}
\cline{3-10}
\multicolumn{2}{c|}{}
       &\multicolumn{4}{c|}{with CEP}&\multicolumn{4}{c|}{without CEP}\\
\hline
nodes &  edges & comp.&clauses&vars&solve&    comp.&clauses&vars&solve \\
\hline
15 & 26 & 0.24 & 13421 & 2154 & 0.07 &
          0.10 & 23424 & 3321 & 0.08 \\

16 & 28 & 0.26 & 18339 & 2851 & 0.19 &
          0.12 & 30136 & 4328 & 0.34 \\

17 & 31 & 0.39 & 21495 & 3233 & 0.07 &
          0.16 & 37074 & 5125 & 0.12 \\

18 & 34 & 0.49 & 26765 & 3928 & 0.12 &
     0.21 & 45498 & 6070 & 0.13 \\

19 & 38 & 0.46 & 30626 & 4380 & 0.11 &
     0.22 & 54918 & 7024 & 0.15 \\

20 & 41 & 0.55 & 43336 & 6005 & 5.93 &
     0.25 & 68225 & 8507 & 12.70 \\

21 & 44 & 0.77 & 52187 & 7039 & 1.46 &
     0.31 & 81388 & 9835 &69.46 \\

22 & 47 & 0.88 & 61611 & 8118 & 71.73 &
     0.35 & 96214 &11276 & 45.43 \\

23 & 50 & 1.10 & 73147 & 9352 & 35.35 &
     0.38 &113180 &13101 & 27.54 \\

24 & 54 & 2.02 & 81634 &10169 & 96.11 &
     0.50 &130954 &14712 & 282.99 \\

25 & 57 & 1.40 & 99027 &12109 & 438.91 &
     0.53 &152805 &16706 & 79.11 \\

26 & 61 & 4.58 &110240 &13143 & 217.72 &
     0.73 &175359 &18615 & 815.55 \\

27 & 65 & 2.16 &127230 &14856 & 35.36 &
     0.75 &201228 &20791 & 114.55 \\
\hline
\end{tabular}
  \caption{Search for graphs with no
    cycles of size 4 or less (comp. \& solve times in sec.)}
  \label{table:cep}
\end{table}



\section{Enhancing \bee\ for Binary Number Representation}

This section describes an extension of \bee\ to support binary
numbers.  A naive extension is straightforward. There is a wide body
of research specifying the bit-blasting of finite domain constraints
for binary arithmetic. So, that is not the topic of this section.  The
interesting aspect of this exercise is how to obtain the constraint
encodings together with support for equi-propagation on their bit
representations. In the presentation we refer to the current version
of \bee\ as the \emph{unary core}, and to the extension for binary
numbers as the \emph{binary extension}.  There are several
possible approaches to define the binary extension:
\begin{enumerate}
\item \underline{CEP}: A straightforward approach is to specify
  standard encodings for each of the new constraints in the binary
  extension and then to flag each of them (individually) as candidates
  for complete equi-propagation. In this way, as described in the
  previous section, \bee\ will infer at compile time all
  equi-propagations and perform the corresponding simplifications.
  However, the implementation of CEP involves calling a SAT solver and
  its application should be limited.

\item \underline{Ad-hoc rules}: Another option is to introduce ad-hoc
  equi-propagation rules for each binary constraint similar to those
  already in \bee\ for the unary constraints (recall the example of
  Figure~\ref{fig:rules2}).
  However, besides being tedious, for the constraints of the binary
  extension there are very few relevant ad-hoc rules.

\item \underline{Decomposition to the unary kernel}: In this approach
  we design encodings for binary constraints in terms of
  decompositions to unary constraints for which equi-propagation rules
  are already defined.
  For example, encoding the multiplication of two $n$-bit binary
  numbers decomposes to involve unary sums of at most $2n$ bits
  each. The unary core then performs equi-propagation on the
  decomposed constraints.

\end{enumerate}

We describe encodings using the third approach for two constraints on
binary representations: \texttt{binary\_array\_sum\_eq} and
\texttt{binary\_times}. 
We also consider the special case where multiplication is applied to
specify that $\mathtt{Z=X^2}$ and demonstrate ad-hoc rules for that
case.

\paragraph{\bf Summing:~~}

Consider a constraint $\mathtt{binary\_array\_sum\_eq(As,Sum)}$ where
$\mathtt{As}$ is an array of binary numbers and $\mathtt{Sum}$ is the
binary number representing their sum.  
In this context, we view \texttt{As} as a binary matrix. The rows
correspond to binary numbers, and the columns to so-called buckets
which are sets of bits with the same ``weight'' or position.
The number of rows is typically not large so that it is reasonable to
sum the columns using unary arithmetic. In this way the decomposition
of the constraint on binary numbers relies on the underlying unary
core of \bee.
Assume that $\mathtt{As}$ consists of more than a single number,
otherwise the decomposition is trivial. The decomposition proceeds as
follows: 
\textbf{(1)} apply $\mathtt{transpose(As,Bs)}$ which transposes the
binary numbers in $\mathtt{As}$ to a bucket representation
$\mathtt{Bs}$ (assume least significant bucket first). \textbf{(2)}
introduce unary-core constraints
$\mathtt{bool\_array\_sum\_eq(B_i,U_i)}$ which sum the buckets to an
array $\mathtt{Us}$ of unary numbers. \textbf{(3)} the recursively
defined $\mathtt{buckets2binary([U|Us],C,[S|Sum])}$ finishes the task
and is defined as follows. 
\[\small \fbox{$\mathtt{buckets2binary([U|Us],C,[B|Sum])}$}
    \xRightarrow[]{\mathtt{decompose}}
    \fbox{$
  \begin{array}{l}
    \mathtt{int\_plus(U,C,U')},\\
    \mathtt{int\_div(U',2,C')},\\
    \mathtt{int\_mod(U',2,B)}, \\
    \mathtt{buckets2binary(Us,C',Sum)}
  \end{array}$}
\]
Here: \texttt{U} is the least significant (unary) bucket, \texttt{C}
is a carry variable (unary integer, initially 0), and \texttt{B} is the
least significant bit of the  (binary) sum.
When, eventually, the buckets are exhausted, decomposition proceeds
as follows.
\[\small \fbox{$\mathtt{buckets2binary([~],C,[B|Sum])}$}
    \xRightarrow[\mathtt{C>0}]{\mathtt{decompose}}
    \fbox{$
  \begin{array}{l}
    \mathtt{int\_div(C,2,C')},\\
    \mathtt{int\_mod(C,2,B)}, \\
    \mathtt{buckets2binary([~],C',Sum)}
  \end{array}$}
\]
Observe that, if applied without any buckets,
$\mathtt{buckets2binary([~],Unary,Binary)}$ defines the channeling
between unary and binary representations.
We also note that for unary numbers, the encoding of division and
modulo by 2 are efficient. Division (by 2) simply collects the even
positioned bits, and modulo (2) takes advantage of the fact that the
representation is ``sorted''.


Below we evaluate our proposed encoding in \bee, but first let us
introduce the encoding of binary multiplication.

\paragraph{\bf Multiplying:~~}

Consider a constraint $\mathtt{binary\_times(A,B,C)}$ specifying that
$\mathtt{C=A\times B}$. It is implemented in \bee\ as follows.  Assume
that $\mathtt{A=[A_n\ldots A_1]}$ and $\mathtt{B=[B_m\ldots B_1]}$ are
the binary representations of A and B.  Decomposition for this
constraint introduces clauses defining
\begin{equation}
  \label{eq:sq}
  \bigwedge_{\tiny{
  \begin{array}{l}
    1\leq i\leq n\\  1\leq j\leq m
  \end{array}}
  } Z_{ij}\leftrightarrow A_i\wedge B_j
\end{equation}
and an additional constraint
$
\mathtt{binary\_array\_sum\_eq([Z_1,\ldots Z_m],C)}
$
where for $1\leq j\leq m$, $Z_j$ is the binary number with
bits \[Z_{nj}\ldots Z_{1j} \underbrace{0\ldots 0}_{j-1}\]
The decomposition is illustrated in
Figure~\ref{fig:square}\subref{fig:square1} where rows 3--7 (with the
$z_{ij}$ variables) are binary numbers to be summed. The encoding
focuses on the corresponding columns which are then encoded to sums
(and carries) using the unary core of \bee\ as described above.

\begin{figure}[t]
  \centering
\subfloat[][Binary multiplication reduces to a sum.]{
$
\begin{array}{@{\extracolsep{-.5mm}}ccccccccccc}
  & & & & & &x_4 &x_3 &x_2 &x_1 &x_0 \\
  & & & & &\mathbf{\times} &y_4 &y_3 &y_2 &y_1 &y_0 \\
\cline{6-11}
  & & & & & &z_{04} &z_{03} &z_{02} &z_{01} &z_{00} \\
  & & & & &z_{14} &z_{13} &z_{12} &z_{11} &z_{10} & \\
  & & & &z_{24} &z_{23} &z_{22} &z_{21} &z_{20} & & \\
  & & &z_{34} &z_{33} &z_{32} &z_{31} &z_{30} & & & \\
  &\mathbf{+} &z_{44} &z_{43} &z_{42} &z_{41} &z_{40} & & & & \\
\hline 
&&&&&&&&&&\\
\end{array}$
\label{fig:square1}}
$~~\xrightarrow[\mathtt{equi.p}]{\mathtt{phase 1}}$
\subfloat[][When $\tuple{x_4,x_3,x_2,x_1,x_0}=\tuple{y_4,y_3,y_2,y_1,y_0}$,
  application of $z_{ij}=z_{ji}$ in bold.]{
$\begin{array}{@{\extracolsep{-.5mm}}ccccccccccc}
  & & & & & &x_4 &x_3 &x_2 &x_1 &x_0 \\
  & & & & &\mathbf{\times} &x_4 &x_3 &x_2 &x_1 &x_0 \\
\cline{6-11}
  & & & & & &z_{04} &z_{03} &z_{02} &z_{01} &z_{00} \\
  & & & & &z_{14} &z_{13} &z_{12} &z_{11} &{\bf z_{01}} & \\
  & & & &z_{24} &z_{23} &z_{22} &{\bf z_{12}} &{\bf z_{02}} & & \\
  & & &z_{34} &z_{33} &{\bf z_{23}} &{\bf z_{13}} &{\bf z_{03}} & & & \\
  &\mathbf{+} &z_{44} &{\bf z_{34}} &{\bf z_{24}} &{\bf z_{14}} &{\bf z_{04}} & & & & \\
\hline
&&&&&&&&&&\\
\end{array}$
\label{fig:square2}}

\subfloat[][Let the bits in each column float down.]{
$\begin{array}{@{\extracolsep{-.5mm}}ccccccccccc}
  &          &     &     &      &      &z_{04} &    &      & & \\
  &          &     &     &      &z_{14}&z_{13} &z_{03}&     & & \\
  &          &     &     &z_{24} &z_{23}&z_{22} &z_{12}&\fbox{$z_{02}$}& & \\
  &          &     &z_{34}&z_{33}&z_{23}&z_{13} &z_{12}&z_{11}&z_{01} & \\
  &\mathbf{+}&z_{44}&z_{34}&z_{24}&z_{14}&z_{04}&z_{03}&\fbox{$z_{02}$}&z_{01} &z_{00} \\
\hline
&&&&&&&&&&\\
\end{array}$
\label{fig:square3}}
$\xrightarrow[\mathtt{equi.p}]{\mathtt{phase 2}}$
\subfloat[][Double bits turn single and move left.]{
$\begin{array}{@{\extracolsep{-.5mm}}ccccccccccc}
  &          &     &     &      &      &      &    &      &      & \\
  &          &     &     &      &      &      &    &     &     & \\
  &          &     &     &z_{23}&      &z_{12} &    &     &      & \\
  &          &z_{34}&     &z_{14}&z_{13}&z_{03} &     &z_{01}&  & \\
  &\mathbf{+}&z_{44}&z_{24}&z_{33}&z_{04}&z_{22} &\fbox{$z_{02}$}&z_{11}& ~0~ &z_{00} \\
\hline
&&&&&&&&&&\\
\end{array}$
\label{fig:square4}}

  \caption{Decomposing the multiplication for the case of a square}
  \label{fig:square}
\end{figure}

To evaluate the encodings of $\mathtt{binary\_array\_sum\_eq(As,Sum)}$
and \\ $\mathtt{binary\_times(A,B,C)}$, we consider the application of
\bee\ to model and solve the $n$-fractions problem, also known as
\texttt{CSPLIB 041}.\footnote{See
  \url{http://www.cs.st-andrews.ac.uk/~ianm/CSPLib/prob/prob041/index.html}}
Here, one should find digit values $(1-9)$ for the variables in
\[\sum_{i=1}^n {{x_i}\over{10*y_i+z_i}}=1\] such that each digit value
is used between 1  and $\lceil{n/3}\rceil$ times. 
Table~\ref{tab:fractions} depicts experimental results comparing two
encodings of $\mathtt{binary\_array\_sum\_eq(As,Sum)}$: Both
techniques sum the columns in the matrix \texttt{As} (where the rows
are binary numbers). The \emph{binary approach} repeatedly reduces
triplets of bits in a column to a pair of bits: one in the same column
(the sum bit), and one in the next (the carry bit). This is a standard
``$3\times 2$'' reduction. The alternative, \emph{unary approach} is
defined in terms of the unary core of \bee.
One may note that the unary approach typically: gives slightly slower
compilation times (there is more to optimize), smaller encoding sizes
(equi-propagation kicks in), and significantly faster SAT solving
times (it pays off) (see footnote [\ref{machine}] for details on
machine).

\begin{table}[t]\small
  \centering
  \begin{tabular}{|l|rrrr|rrrr|}
\hline
           & \multicolumn{4}{c|}{summing with full adders} 
           & \multicolumn{4}{c|}{summing in the unary core}\\
\hline
 $n$      &comp.&clauses&vars&sat&comp.&clauses&vars&sat\\
\hline
  3 & 0.05&25492 & 4354&2.72   & 0.26&23793 &4556 &1.39\\
  4 & 0.13&56125 & 9556&11.19  & 0.50&47743 &9078 &0.56\\
  5 & 0.23&98712 &16551&59.4   & 0.77&78607 &14703&55.65\\
  6 & 0.38&164908&27283&844.91 & 1.01&118850&21977&5.13\\
  7 & 0.76&247082&40572&$\infty$~~ & 1.87&164451&30125&36.83\\
  8 & 1.29&363323&59183&$\infty$~~~& 2.14&221262&40196&2653.68\\
\hline
\end{tabular}

\caption{Comparison of encodings for the $n$-fractions problem
  (comp. and sat times in sec. with 4 hour timeout marked as $\infty$)}
\label{tab:fractions}
\end{table}

\paragraph{\bf Squaring:~~}

Consider the special case of multiplication
$\mathtt{binary\_times(A,A,C)}$ specifying that $\mathtt{A^2=C}$ where
we introduce two additional optimizations. First, consider the
variables $\mathtt{z_{ij}}$ introduced in Equation~\ref{eq:sq}, we
have $Z_{ij}=Z_{ji}$ and hence equi-propagation applies to remove the
redundant variables. The result of this is illustrated in
Figure~\ref{fig:square}\subref{fig:square2}. In
Figure~\ref{fig:square}\subref{fig:square3} we reorder the bits in the
columns, as if, letting the bits drop down to the baseline.  Second,
consider the ``columns'' in the
$\mathtt{binary\_array\_sum\_eq([Z_1,\ldots Z_m],C)}$ constraint. Each
variable of the form $Z_{ij}$ with $i\neq j$ in a column occurs
twice. So, both can be removed and one inserted back in the column to
the left. This is illustrated in
Figure~\ref{fig:square}\subref{fig:square4} where we highlight the
move of the two $z_{02}$ instances. These optimizations reduce the
size of the CNF and are applied both in the binary and in the unary
encodings.

To evaluate the encoding of $\mathtt{binary\_times(A,B,C)}$ for the
special case when $\mathtt{A=B}$, we consider the application of \bee\
to model and solve the number partitioning problem, also known as
\texttt{CSPLIB 049}.\footnote{See
  \url{http://www.cs.st-andrews.ac.uk/~ianm/CSPLib/prob/prob049/index.html}}
Here, one should finding a partition of numbers $\{1,\ldots,n\}$ into
two sets $A$ and $B$ such that: $A$ and $B$ have the same cardinality,
the sum of numbers in $A$  equals the sum of numbers in $B$, and
the sum of the squares of the numbers in $A$ equals the sum of the
squares of the numbers in $B$. 

\begin{figure}
  \centering
  
  \subfloat[][Encoding sizes (number of variables): unary and binary
              approach, with and without CEP]{
\includegraphics[width=0.45\linewidth,keepaspectratio]{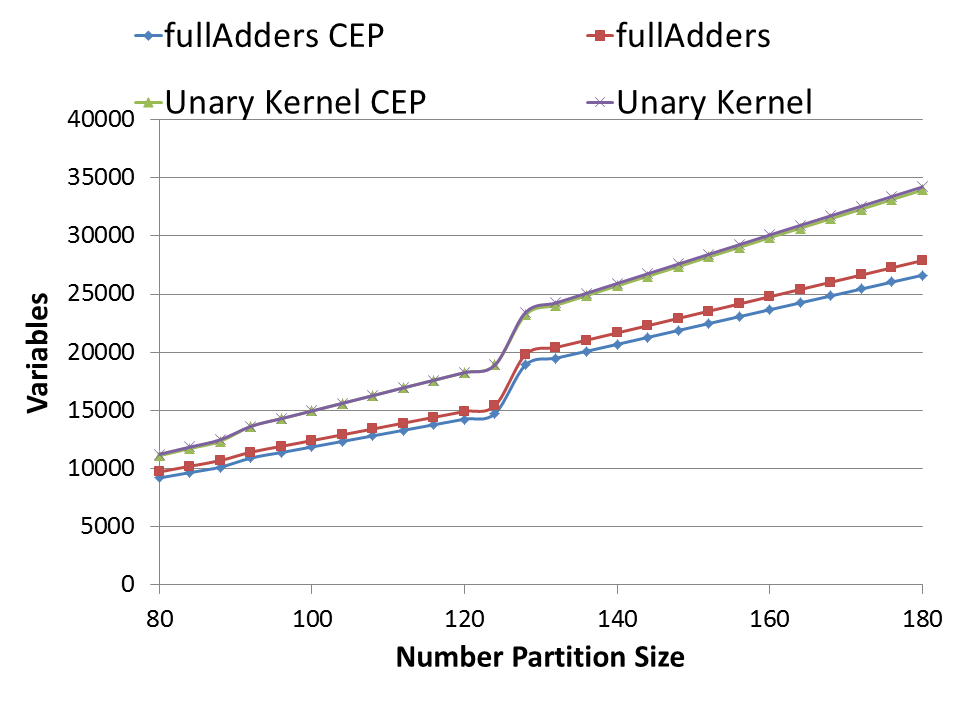}
\label{tab1}}
\subfloat[][SAT solving  time (sec.)]{
\includegraphics[width=0.45\linewidth,keepaspectratio]{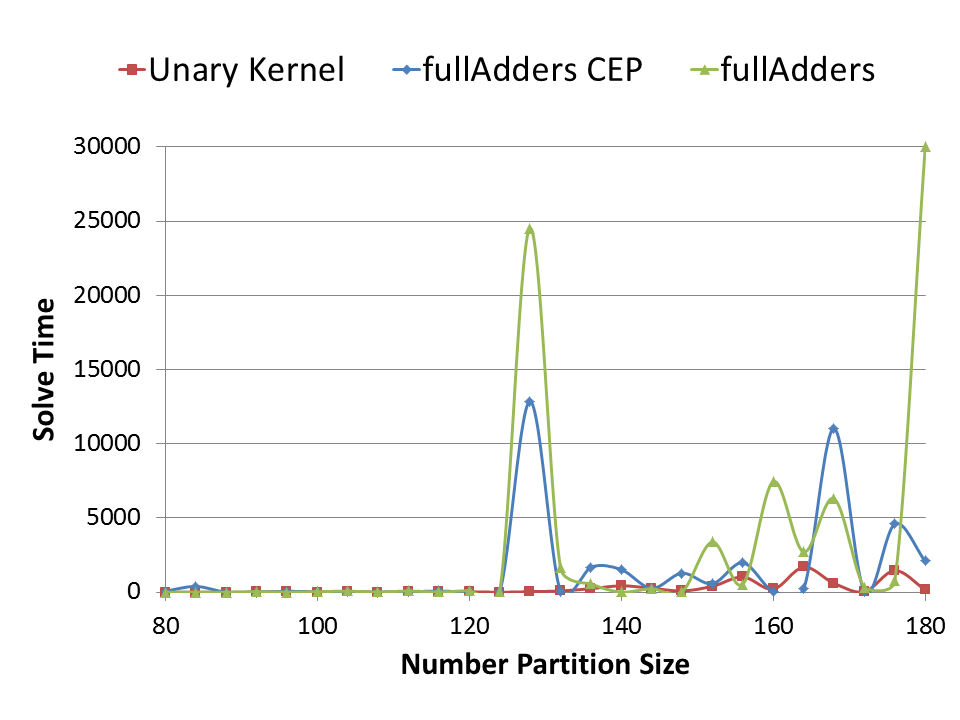}
\label{tab2}}

\subfloat[][Encoding size, CEP minus without (\# clauses)]{
\includegraphics[width=0.45\linewidth,keepaspectratio]{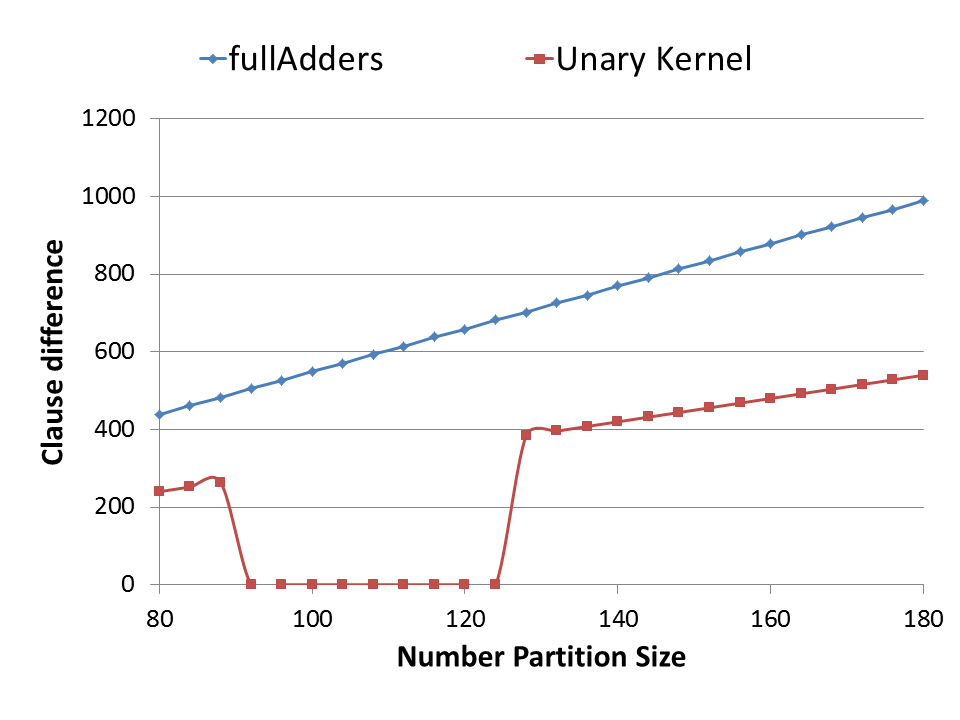}
\label{tab3}}
\subfloat[][Encoding size, CEP minus without (\# vars)]{
\includegraphics[width=0.45\linewidth,keepaspectratio]{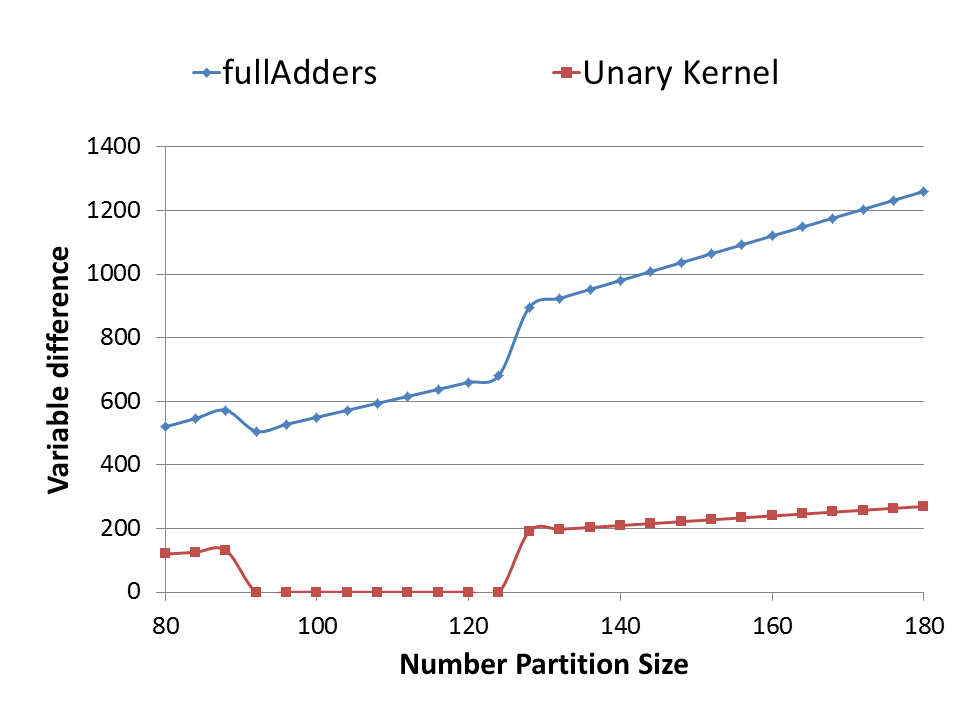}
\label{tab4}}

\caption{Number Partitioning in \bee, encoding binary arithmetic.}
\label{fig:compareNumPartition}
\end{figure}
 
Figure \ref{fig:compareNumPartition} depicts our results. We consider
four settings. The first two are the binary and unary approaches
described above where buckets of bits of the same binary weight are
summed using full adders or sorting networks respectively. In the
other two settings, we apply complete equi-propagation per individual
constraint (on binary numbers), on top of the ad-hoc rules implemented
in \bee.
Figure~\ref{fig:compareNumPartition}\subref{tab1} illustrates the size
of the encodings (number of CNF variables) for each of the four
settings in terms of the instance size. The two top curves coincide
and correspond to the unary encodings which create slightly larger
CNFs. However note that the unary core of \bee\ with its ad-hoc (and
more efficient) implementation of equi-propagation, detects all of the
available equi-propagation. There is no need to apply CEP. The bottom
two curves correspond to the binary encodings and illustrate that CEP
detects further optimizations beyond what is detected using \bee.

Figure~\ref{fig:compareNumPartition}\subref{tab2} details the SAT
solving times. Here we ignore the compilation times (which are high
when using CEP) and focus on the quality of the obtained CNF. The
graph indicates a clear advantage to the unary approach (where CEP
is not even required).
The average solving time using the unary approach approach (without
CEP) is 270 (sec) vs 1503 (sec) using the binary approach (with
CEP). This is in spite of the fact that unary approach involves larger
CNF sizes.

Figures~\ref{fig:compareNumPartition}\subref{tab3} and \subref{tab4}
further detail the effect of CEP in the binary and unary encodings
depicting the numbers of clauses and of variables reduced by CEP in
both techniques. The smaller this number, the more equi-propagation
performed ad-hoc by \bee.
In both graphs the lower curve corresponds to the encodings based on
the unary core indicating that this is the one of better quality. See
footnote [\ref{machine}] for details on machine.

\section{Conclusion}

We have detailed two features of \bee\ not described in previous
publications. These concern the hybrid approach to encode cardinality
constraints and the procedure for applying complete
equi-propagation. We have also described our approach to enhance the
unary kernel of \bee\ for binary numbers. Our approach is to rely as
much as possible on the implementation of equi-propagation on unary
numbers to build the task of equi-propagation for binary numbers. We
have illustrated the power of this approach when encoding binary
number multiplication. The extension of \bee\ for binary numbers is
ongoing and still requires a thorough experimentation to evaluate its
design.

\newpage

\newcommand{\noopsort}[1]{}

\end{document}